\documentclass{emulateapj}

\newcommand{\myemail}{sascha.quanz@astro.phys.ethz.ch}

\slugcomment{Accepted version - \\To appear in a revised / edited form in the International Journal of Astrobiology published by Cambridge University Press}

\shorttitle{Direct detection of exoplanets in the 3 -- 10 micron range with E-ELT/METIS}
\shortauthors{Quanz et al.}

\begin{document}


\title{Direct detection of exoplanets in the 3 -- 10 micron range with E-ELT/METIS}

\author{Sascha P. Quanz$^{1}$, Ian Crossfield$^2$, Michael R. Meyer$^1$, Eva Schmalzl$^3$, Jenny Held$^{4}$}
\email{\myemail}

\altaffiltext{1}{Institute for Astronomy, ETH Zurich, Wolfgang-Pauli-Strasse 27, 8093 Zurich, Switzerland}
\altaffiltext{2}{Max Planck Institute for Astronomy, Ko\"nigstuhl 17, 69117 Heidelberg, Germany}
\altaffiltext{3}{Leiden Observatory, Leiden University, P.O. Box 9513, 2300 RA Leiden, The Netherlands}
\altaffiltext{4}{Department of Physics, ETH Zurich, 8093 Zurich, Switzerland}

\begin{abstract}
We quantify the scientific potential for exoplanet imaging with the Mid-infrared E-ELT Imager and Spectrograph (METIS) foreseen as one of the instruments of the European Extremely Large Telescope (E-ELT). We focus on two main science cases: (1) the direct detection of known gas giant planets found by radial velocity (RV) searches; and (2) the direct detection of small (1 -- 4 R$_\earth$) planets around the nearest stars. Under the assumptions made in our modeling, in particular on the achievable inner working angle and sensitivity, our analyses reveal that within a reasonable amount of observing time METIS is able to image $>$20 already known, RV-detected planets in at least one filter. Many more suitable planets with dynamically determined masses are expected to be found in the coming years with the continuation of RV-surveys and the results from the GAIA astrometry mission. In addition, by extrapolating the statistics for close-in planets found by \emph{Kepler}, we expect METIS might detect $\approx$10 small planets with equilibrium temperatures between 200 -- 500 K around the nearest stars. This means that (1) METIS will help constrain atmospheric models for gas giant planets by determining for a sizable sample their luminosity, temperature and orbital inclination; and (2) METIS might be the first instrument to image a nearby (super-)Earth-sized planet with an equilibrium temperature near that expected to enable liquid water on a planet surface.
\end{abstract}


%% subject header.  Objects should be in the appropriate "individual"
%% headers (e.g. quasars: individual, stars: individual, etc.) with the
%% additional provision that the total number of headers, including each
%% individual object, not exceed six.  The \objectname{} macro, and its
%% alias \object{}, is used to mark each object.  The macro takes the object
%% name as its primary argument.  This name will appear in the paper
%% and serve as the link's anchor in the electronic edition if the name
%% is recognized by the data centers.  The macro also takes an optional
%% argument in parentheses in cases where the data center identification
%% differs from what is to be printed in the paper.


\section{Introduction}
Since the discovery of the first exoplanet orbiting a Sun-like star in 1995 \citep{mayorqueloz1995} radial velocity (RV) measurements and transit photometry have been the dominate techniques for the detection and characterization of exoplanets. 
Long-term ground-based RV surveys and in particular NASA's space mission \emph{Kepler} enable us to constrain the occurrence rate of planets around other stars as a function of planetary mass, size, and orbital period. \citet{mayor2011} presented RV statistics for Solar-type host stars and planets with periods out to 10 years and minimum masses as low as $\sim$3 -- 10 M$_\earth$. Similarly, \citet{howard2012} analyzed the occurrence rate of planets with sizes $>$2 R$_\earth$ and orbital periods $\leq 50$ days ($\sim$0.25 AU) around Solar-type stars using \emph{Kepler} data. Both studies demonstrated that low-mass / small planets are much more frequent than massive / large planets. 

As the probability of observing a planet transiting in front of its host stars decreases with orbital separation and as the detection of a planet by RV measurements normally requires the coverage of a full orbit, these two techniques typically reveal planets in the inner few AU around a star. This means that in order to get a census of exoplanets at larger orbital separations other techniques have to complement RV and transit searches. The two most important techniques here are microlensing \citep[e.g.,][]{cassan2012} and direct imaging. We will focus on the latter one throughout this article.

Numerous direct imaging surveys carried out in the last years rule out the existence of a large population of massive gas giants planets ($\gtrsim$2 M$_{\rm Jupiter}$) on wide orbits ($\gtrsim$50 AU) \citep[e.g.,][]{lafreniere2007,chauvin2010,heinze2010_2,rameau2013_2,biller2013,nielsen2013,janson2013,wahhaj2013}\footnote{We only list surveys with $\geq$50 stars.}. However, a few fascinating systems were revealed by direct imaging \citep{marois2008,marois2010,kalas2008,lagrange2010,rameau2013,carson2013,kuzuhara2013} partly challenging our understanding of the formation and atmospheric properties of gas giant planets. The key challenge for direct imaging is to obtain a high-contrast performance at very small inner working angles (IWA) so that the faint signal of a planetary companion can be detected and separated from the strong signal of the nearby star. If successful, the direct detection of photons from an exoplanet offers a unique pathway to study and characterize the planet's atmospheric properties. Transit or secondary eclipse photometry and spectroscopy can in principle also be used to probe the atmosphere of exoplanets directly and currently these techniques produce most of the corresponding results \citep[e.g.,][]{seagerdeming2010}. However, typically only close-in planets can be studied in this way because the probability of transit declines with increasing semimajor axis. As the orbital inclination of exoplanets is randomly distributed in the sky when seen from Earth, the vast majority of the overall exoplanet population does \emph{not} transit in front of and behind their host star. In order to investigate these objects and study exoplanet atmospheres covering a wide range of planetary masses and orbital separations, direct imaging observations - ground-based or space-based - are essential.

Currently, dedicated high-contrast exoplanet imagers are being installed at 8-m telescopes: SPHERE at the VLT (Very Large Telescope) \citep{beuzit2006} and GPI on Gemini South \citep{macintosh2006}. These instruments work in the optical and/or near-infrared and, depending on their final on-sky performance, they are expected to detect gas giant planets down to 10 AU or so around young, nearby stars. Smaller and shorter period planets are typically beyond the reach of these instruments as their contrast performance, sensitivity and spatial resolution are still insufficient\footnote{The ZIMPOL sub-instrument of SPHERE should be able to detect polarized, reflected light of close-in gas giant planets at optical wavelengths around the nearest stars \citep{schmid2006}.}. 

In this paper, we quantify two exoplanet imaging science cases for ground-based Extremely Large Telescopes (ELTs). Our focus is on the 3 -- 10 $\mu$m wavelength range, which is normally dominated by thermal emission from the planets rather than by reflected starlight. We take METIS \citep{brandl2012} as our default instrument in our analyses. METIS could be the third instrument installed at the 39-m European Extremely Large Telescope (E-ELT) according to the E-ELT instrumentation roadmap\footnote{see, http://www.eso.org/sci/facilities/eelt/instrumentation/}. It is AO-assisted and currently foreseen to offer imaging and medium-resolution spectroscopy over the full L, M and N band wavelength range (3 -- 14 microns). High-resolution integral field spectroscopy is planned for the L and M bands (3-5.3 microns). In section 2 we motivate the use of this wavelength range for exoplanet research. In section 3 we quantify two major exoplanet science cases for METIS: (a) the fraction of the currently known exoplanets detected by radial velocity that can be imaged with E-ELT/METIS; and (b) the prospects of directly detecting small planets around nearby stars. We discuss our findings in section 4 and conclude in section 5.

\section{Motivation for thermal infrared imaging of exoplanets from the ground}
The motivation to exploit thermal infrared (IR) wavelengths\footnote{Throughout this paper we use the term thermal IR for the 3 -- 10 $\mu$m wavelength range while near-infrared (NIR) refers to 1 -- 2.5 $\mu$m.} for exoplanet imaging has a scientific and a technical aspect. 
\subsection{Scientific considerations}
By observing at thermal IR wavelengths a slightly different part of the exoplanet parameter space is probed compared to current observations in the NIR. This becomes clear just by considering Wien's law and estimating the blackbody temperatures that correspond to the central wavelength of each filter. For the direct detection of thermal emission from self-luminous gas giant exoplanets, for a given age, one is able to probe less massive planets or, for a given mass, one is able to search around older stars. This is a direct consequence of the very red infrared colors of planetary mass objects and the fact that they contract and cool during their evolution \citep[for theoretical work on gas giant planet evolution and luminosities see, e.g.,][]{burrows2001,chabrier2000,baraffe2003,sudarsky2003,marley2007,fortney2008,spiegel2012}. A nice example, demonstrating the power of observing in the thermal infrared, is provided by the recently discovered planetary mass companion to the A-type star HD95086 by \citet{rameau2013}. While the planet was clearly detected in the L band (3.8 $\mu$m) it remained undetected in H and Ks \citep[$\sim$1.6 and 2.3 $\mu$m;][]{meshkat2013,rameau2013}.

Another aspect is the search for planets that are still somewhat embedded in the circumstellar disks of their host stars. The opacities of dust grains typically found in these disks have wavelength dependent extinction effects, with shorter wavelengths photons being more strongly affected. Hence, it could be possible to detect young planets in the thermal infrared that remain unseen at NIR wavelengths. A possible example is the candidate protoplanet detected in the disk around the young intermediate mass star HD100546. \citet{quanz2013} detected the object in the L band. If all of the observed flux arose from the photosphere of a "normal" low-mass companion its mass would be 20 -- 25 M$_{\rm Jupiter}$ according to theoretical models. \citet{boccaletti2013} analyzed K band data from Gemini/NICI of the same star and were not able to detect the companion candidate even though the data were good enough to see a 15 -- 20 M$_{\rm Jupiter}$ object based on the same models. In the meantime, the object was re-detected in a new L band dataset (Quanz et al., in prep.). Whether extinction is the dominant effect here or whether the intrinsic properties of the object only allow for a detection in the L band is still to be investigated, but it clearly shows that observations at thermal infrared wavelengths can reveal objects that are not easily accessible with NIR observations. 

For old exoplanets, depending on the instrument performance and the properties of the exoplanetary system, it might be "easier" to detect reflected starlight from the planet at optical or NIR wavelengths than thermal emission at longer wavelengths. However, typically thermal emission from planets depends significantly less on the orbital phase and orbital inclination of the object compared to observations in reflected light. Also, reflected light observations give only access to the product of atmospheric albedo and planet radius and additional observations are required to break the degeneracy. Having an estimate for the effective temperature of the planet from its thermal emission and knowing the distance to the object directly yields its radius. We emphasize that in the ideal case one wants to combine the information obtained from thermal emission \emph{with} those from reflected light. This provides complementary insights in atmospheric properties and allows for a significantly higher degree of characterization \citep[see, e.g.,][]{seager2013}.

Finally, once a planet has been detected, it is important to understand the diagnostic power of different wavelength bands for the characterization of the object. Concerning potential atmospheric constituents is worth recalling that the L, M and N bands include some main molecular features, for example CH$_4$ (3.3  $\mu$m), CO (4.7 $\mu$m), and O$_3$ (4.7 $\mu$m, 9.6  $\mu$m). In particular for the gas giant planets that have already been directly imaged (e.g., the HR8799 system) the spectral energy distribution (SED) in the L band between 3.2 -- 4.0 $\mu$m is of substantial diagnostic power to constrain the ratio between CO and CH$_4$ in the atmosphere and to search for potential chemical disequilibrium \cite[e.g.,][]{hinz2010,skemer2012}. Additional indications for chemical disequilibrium could also come from the very red end of the L band. \citet{janson2011} took a low-resolution L band spectrum of the exoplanet HR8799 c and found an apparent deficiency in flux beyond 4 $\mu$m as predicted by non-equilibrium atmospheric models. In addition, clouds seem to play a crucial role in shaping the SEDs of these massive planets. 
Also here, the L band regime proves to be an important diagnostic window \citep[e.g.,][]{lee2013}. 
Turning to smaller, rocky planets it turns out that the N band could potentially be used to constrain the surface composition of warm/hot objects. \citet{hu2012} presented simulations of 8 -- 13 $\mu$m spectra for rocky planet with different surface temperatures and compositions. Surface characterization could provide a powerful method to unambiguously identify a rocky, airless exoplanet.

\subsection{Technical considerations}
As mentioned above, the direct detection and characterization of exoplanets requires high-contrast performance at very small IWA. Given their very different effective temperatures, the flux contrast between stars and planets in the 3 -- 10 $\mu$m range is less stark than at optical or near-infrared wavelengths. Hence, the instrument requirements in terms of achievable contrast performance can be less stringent for observation in the thermal infrared. In addition, AO systems on large ground-based telescopes provide higher Strehl ratios at thermal infrared wavelengths compared to the optical or NIR. This, again, helps to tackle the contrast problem as more light from the central star is concentrated in the core of the PSF and less flux is left in the uncorrected halo. AO-assisted thermal infrared imagers provide Strehl ratios in the L and M band that are at least comparable, if not superior, to the expected Strehl ratios of the next generation high-contrast imagers working in the NIR. LMIRCam at the Large Binocular Telescope (LBT) is equipped with an adaptive secondary mirror and typically reaches Strehl ratios of 80 -- 90\% in the L and M band. This is also the goal for the planned ERIS instrument at the VLT. 

The key challenge to overcome, when observing in the thermal infrared from the ground, is the background emission from the sky, the telescope and the optical components within the instrument. On the one hand, the photon noise from these contributions sets the ultimate detection limit for a given observing time, but temporal fluctuations in the background pose an additional problem. Great improvements can be made if this issue is taken into account already in the design phase of an instrument by minimizing the amount of "warm" components in the light path. Good examples are CLIO, formerly installed at the Multiple Mirror Telescope (MMT), LBT/LMIRCam and VLT/ERIS.

A technical goal for high contrast imaging at thermal infrared wavelengths is to move the background limited regime as close as possible to the central star. If one is background-limited and not contrast-limited then - to first order - by increasing the observing time, higher sensitivities can be reached (the SNR is then proportional to $\sqrt t $ with $t$ being the integration time). Different coronagraphic designs to cancel out the diffraction rings around the central star have been developed in recent years specifically for observations between 3 -- 10$\mu$m. Some of them have been installed on 8-m telescopes \citep[e.g., NACO/APP, NACO/AGPM;][]{kenworthy2010,mawet2013}. Typically, these coronagraphs in combination with optimized imaging techniques \citep[e.g., ADI;][]{marois2006} and advanced data reduction algorithms \citep[e.g., PynPoint;][]{amaraquanz2012} aim at moving the background limit (BGL) as close as $2 \lambda_{{\rm cen}}/D$ to the star, which would then correspond to the high-contrast IWA of the imaging system. 

One key aspect we would like to emphasize is that with the advent of 25 -- 40 m diffraction limited telescopes, observations at thermal infrared wavelengths become significantly more efficient. In the sky background limited case and for a given SNR the time to complete an observation scales as $t\propto D^{-4}$, which is an enormous advantage for future extremely large telescopes compared to current 8 m class telescopes.

\begin{deluxetable}{cccccc}
\tablecaption{E-ELT METIS filter and performance estimates.
\label{metis_table}}           % title of Table
\tablewidth{\linewidth}
\tablehead{
\colhead{Filter}  & \colhead{$\lambda_{{\rm cen}}$}   & \colhead{Filter width} &  \colhead{IWA\tablenotemark{a}} & \colhead{Sensitivity\tablenotemark{b}} & \colhead{Limiting}\\
\colhead{}  & \colhead{[$\mu$m]} & \colhead{[$\mu$m]} &  \colhead{$['']$} & \colhead{[$\mu$Jy]} & \colhead{magnitude\tablenotemark{b,c}}
}
\startdata
L & 3.58 & 0.98 & 0.038 & 0.27 & 22.4 \\
M & 4.78 & 0.60 & 0.051 & 2.76 & 19.3 \\
N & 10.6 & 5.2 & 0.112 & 9.84 & --- \\
\enddata
\tablenotetext{a}{We assume to be background limited down to an IWA of $2 \lambda_{\rm cen}/D$.}
\tablenotetext{b}{5-$\sigma$ detection limits in 3 hours of telescope time incl. 20\% overhead, i.e., 8640 seconds effective on source integration time. These values assume a cold-stop in the instrument design that leads to an effective telescope diameter of D = 37 m and has an inner obscuration of 11.1 m. 
}
\tablenotetext{c}{As the exact filter profiles are not defined yet, the zero points were estimated based on www.not.iac.es/instruments/notcam/ReferenceInfo/conver.html and should be good to 5-10\%. As none of our analyses requires a limiting magnitude in the N band, we do not provide an estimate here.}
\end{deluxetable}

\begin{figure*}
\centering
\epsscale{1.1}
\plotone{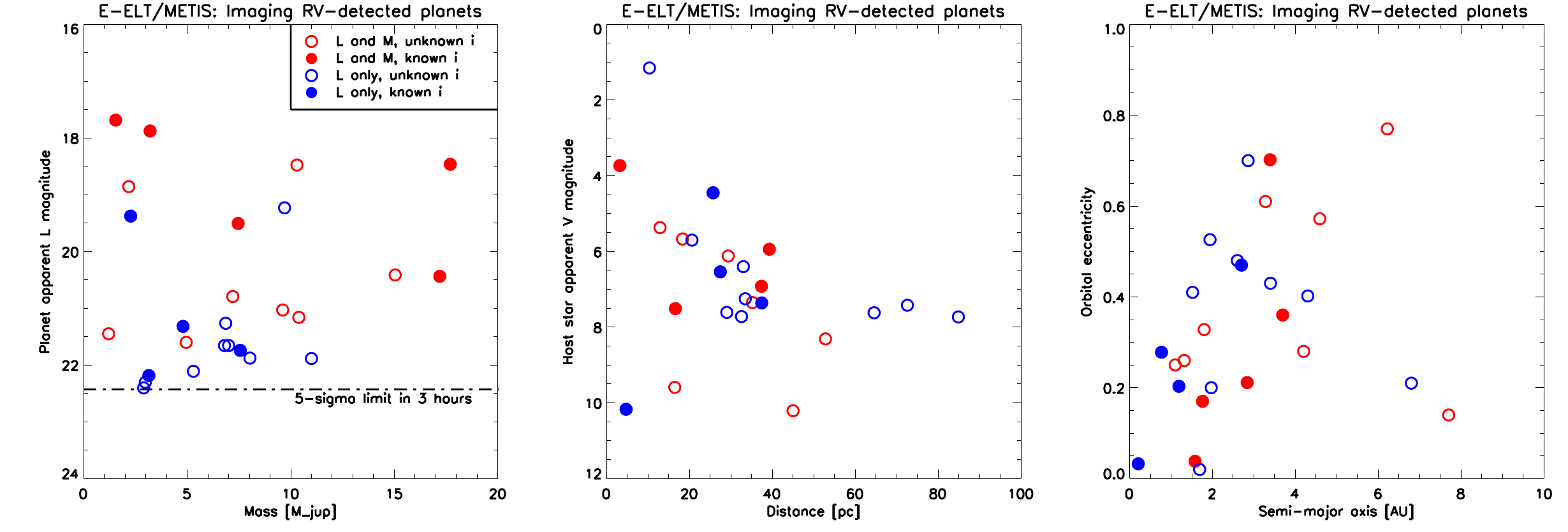}
\caption{Properties of RV-detected planets that can be directly imaged with E-ELT/METIS. Symbols are the same in all panels. Left panel: Apparent L magnitude of planets detected by RV as a function of their (minimum) mass. The dash-dotted line indicates the 5-$\sigma$ detection limit (see, Table~\ref{metis_table}). Blue dots show planets that are only detectable in the L band (13 objects) and red dots planets that are detectable in the L and M band (13 objects). Filled dots are objects with an estimate for their orbital inclination $i$, open dots are objects with unknown $i$. Middle panel: Host star apparent V band magnitude as a function of distance for detectable planets. Right panel: Orbital eccentricity as a function of semi-major axis for detectable planets.
\label{fig.1}}
\end{figure*}

\section{Exoplanet imaging science with E-ELT/METIS}
Two major exoplanet imaging science cases for E-ELT/METIS will be: (a) imaging known exoplanets detected by radial velocity; and (b) directly detecting small planets around nearby stars. Both science cases will be quantified in Sections 3.1. and 3.2., but we emphasize that in addition to these there are other exoplanet science cases that E-ELT/METIS will be able to tackle. These include, for instance, a classical survey for gas-giant and icy planets around the nearest stars; a search for young, forming planets still embedded in the circumstellar disk of their host stars; constraining atmospheric composition of cool gas giant planets \citep[e.g.,][]{janson2011}; or measuring molecular abundances and wind speeds in the atmospheres of hot giant planets using high-resolution spectroscopy \citep[e.g.,][]{snellen2010,brogi2012,birkby2013}.

As mentioned above, for the following analyses we focus on the imaging performance of METIS and do not consider spectroscopic applications. METIS will offer diffraction limited imaging in the L, M and N bands with a field-of-view (FoV) of approximately 18$''\times$18$''$\footnote{The FoV is large enough as not to put any constraints on the planet detections discussed in this paper. The on-sky separation between the planets considered here and their host stars is significantly smaller.}. The assumed sensitivity limits for each band correspond to a three hour observing block. For deep planet searches using the ADI technique such an observing block is typically centered around the targets' meridian passage to maximize the amount of field rotation. As METIS will be situated on Cerro Armazones in Chile, for our final analyses we only consider objects with a declination $<$30$^\circ$, which ensures the airmass is $<$2.0 for all objects. In Table~\ref{metis_table} we summarize the assumed filter properties, the inner working angles and the sensitivity estimates for the different bands. 

\subsection{Direct detection of planets found by radial velocity surveys}
Direct imaging of planets found by radial velocity surveys offers major opportunities. Irrespective of the age of the system, multiple epoch observations yield the orbital inclination of the planet and allow for the derivation of the planet's true mass. This information, together with the observed brightness (possibly in multiple bands) and host star properties, can then be used to characterize individual planets and to constrain atmospheric models of planetary mass objects. At young ages, direct imaging is the only technique that can  break some of the degeneracies existing in the atmospheric and evolutionary models of gas giant planets and to constrain the initial conditions for gas giant planet formation. As these conditions are basically unknown, a huge spread in fundamental planet parameters (e.g., radius and temperature) is found for ages $\lesssim$200 Myr, depending on what initial conditions are chosen \citep[e.g.,][]{spiegel2012}. Up to now, $\beta$ Pic b is the only young planet, where we have images as well as at least some mass constraints derived from radial velocity measurements \citep{lagrange2012}. 

In the following, based on the current census of exoplanets discovered by RV, we quantify how many planets an instrument such as METIS at the E-ELT could detect. To do so, we retrieved the list of known exoplanets detected by RV from the exoplanet.eu database \citep[as of July 29, 2013;][]{schneider2011}. Not all relevant parameters are known for all of these systems. To be conservative we disregarded objects with unknown values for the semi-major axis $a$, orbital period $P$, distance from Earth to the host star $d$, or argument of periastron $\omega$. In cases where the orbital eccentricity $e$ was not known, we conservatively assumed $e=0$. If the orbits were, however, eccentric, the planets would be easier to detect as they would spend more time at larger separations from their host star. For systems with no estimate for the age, we assumed an age of 5 Gyr, which is the average age of the planets with age estimates. Furthermore, we only considered objects with a minimum mass of $m\cdot\sin(i) > 0.3$ M$_{\rm Jupiter}$. Initial estimates suggested that planets with lower minimum masses would not be easily detectable with our underlying assumptions. This is supported by our final results, where the lowest mass planet that is detectable has $m\cdot\sin(i) > 1$ M$_{\rm Jupiter}$ (see below) even though numerous planets with masses 0.3 M$_{\rm Jupiter}$ $< m\cdot\sin(i) < 1.0$ M$_{\rm Jupiter}$ were included in the input sample. In total, the initial input sample had 352 objects, still including objects from both the Northern and Southern hemisphere.

We used the COND atmospheric models \citep{baraffe2003} to compute the apparent L and M band magnitudes for the planets as a function of age and minimum mass. This approach is again conservative in 2 ways: (1) Using the given minimum mass to estimate the planets' brightness is conservative, as planets with higher masses would be brighter and hence easier to detect at all ages; (2) for some planets the expected equilibrium temperature $T_{eq}$ may exceed the effective temperature predicted by the COND models $T_{eff,model}$ in which case the planets' brightness is also underestimated. 

In a next step, we estimated the typical on-sky separation between the planet and its host star and compared it to the assumed IWA of METIS in the L and M band. In cases where the orbital inclination $i$ of the planet is known (or at least estimated), it is straight forward to compute the apparent separation between planet and star as a function of time. For most RV planets, however, $i$ and also the longitude of the ascending node $\Omega$ are unknown. We did a Monte-Carlo simulation for these two variables to estimate the probability distribution of the on-sky separation knowing all the other orbital parameter. We then only kept (1) planets with known $i$ where the apastron passage exceeds a separation of $2 \lambda_{{\rm cen}}/D$, and (2) planets with unknown $i$ and $\Omega$, where the probability of finding the planet at a separation $> 2 \lambda_{{\rm cen}}/D$ exceeds 50\%, if one were to observe the object at a random time. This selection resulted in 130 and 95 objects in the L and M band, respectively. We then excluded 25 objects in the L band and 19 objects in the M band as their declination was higher than our assumed limit.

Finally, we compared the estimated L and M band magnitudes of the remaining objects to our assumed 5-$\sigma$ sensitivity limits (Table~\ref{metis_table}). Figure~\ref{fig.1} shows that, given the selection criteria described above, METIS is capable of directly detecting 26 of the known gas giant planets in the L band, 13 of which are also detectable in the M band. The planets cover a range in minimum mass roughly between 1 and 18 M$_{\rm Jupiter}$ and their host stars span a wide magnitude range for a given distance, which indicates an interesting spread in both planet and also host star properties. Furthermore, the planets span a wide range in semi-major axis and also orbital eccentricity. This leads to a wide range of planetary temperatures across the whole sample, but also to significant changes in stellar insolation for individual planets on highly eccentric orbits. A first-order estimate for the expected equilibrium temperature reveals that 2 planets (HD 62509 b and HD 60532 b) likely have effective temperatures higher than those predicted by the models applied in our selection process. Hence, these planets should appear even brighter and be easier to detect than shown here. Finally, three of the planets reside in stellar binary systems (HD196885 A b, HD106515 A b, and  GJ676 A b), and there are two systems where two planets can be detected (HD 60532 b,c and HD 128311 b,c). These latter systems potentially allow a direct comparison of gas giant properties within extrasolar planetary systems. Other stars have additional planets as well, but those are below the detection limits chosen here. We list all planets and their key properties in Table~\ref{rv_planets}.

\begin{deluxetable*}{lcccccccc}
\tablecaption{Key properties of planets detected by RV surveys that can be imaged with E-ELT/METIS (see text for selection criteria). All values were adopted from the exoplanet.eu database (unless indicated otherwise) except for the apparent L magnitude and the M band flag, which were derived here (see Sec. 3.1.). 
\label{rv_planets}}           % title of Table
\tablewidth{\linewidth}
\tablehead{
\colhead{Name}  & \colhead{Minimum mass}   & \colhead{Apparent L} &  \colhead{Distance} & \colhead{Age} & \colhead{$a$} & \colhead{$e$} & \colhead{Known $i$} & \colhead{M band\tablenotemark{a}} \\
\colhead{}  & \colhead{[M$_{\rm Jupiter}$]}   & \colhead{[mag]} &  \colhead{[pc]} & \colhead{[Gyr]} & \colhead{[AU]} & \colhead{} & \colhead{} & \colhead{}
}
\startdata
%NORTH HD 87883 b&       12.1&      19.71&       18.1&       3.6&      0.53&       yes &      yes  \\
HD 82943 b&       4.8&      21.3&       27.5 &   3.1 &    1.2&      0.20&       yes&      no \\
HD 60532 b&       3.2&      22.2\tablenotemark{b}&       25.7&    2.7 &   0.8&      0.28&       yes&      no \\
HD 60532 c&       7.5&      19.5&       25.7&    2.7 &   1.6&     0.04&       yes&      yes \\
%NORTH ups And d&       10.19 &      17.82&       13.5 &       2.6&      0.27&       yes&      yes \\
eps Eridani b&       1.6&      17.7&       3.2&    0.7 &   3.4 &      0.70&       yes&      yes \\
 HD 168443 c&       17.2&      20.4&   37.4 &   9.8 &    2.8&      0.21 &       yes&      yes \\
HD 38529 c&       17.7&      18.5&       39.3 &  3.3 &     3.7 &      0.36&       yes&      yes \\
HD 128311 c&       3.2&      17.9&       16.6 &  0.4 &     1.8&      0.17&       yes&      yes \\
Gliese 876 b&       2.3&      19.4&       4.7 &     2.5 & 0.2 &     0.03&       yes&      no \\
HD 106252 b&       7.6&      21.7&       37.4 &  5.0 &     2.7&      0.47 &       yes&      no \\
HD 128311 b&       2.2&      18.9&       16.6&   0.4 &    1.1 &      0.25&       no&      yes \\
HD 147513 b&       1.2&      21.5&       12.9&    0.7 &   1.3&      0.26 &       no&      yes \\
%NORTH HD 150706 b&       2.71&      21.54&       27.2 &       6.7&      0.38 &       no&      no \\
%NORTH HD 8673 b&       14.2&      18.14&       38.3 &       3.0&      0.72&       no&      yes \\
HD 196885 A b&       3.0&      22.3&       33.0&  2.0 &     2.6 &      0.48 &       no&      no \\
HD 125612 d&       7.2&      20.8&       52.8 &    2.1&    4.2 &      0.28&       no&      yes \\
HD 86264 b&       7.0&      21.7&        72.6&   2.2 &    2.9&      0.70&       no&      no \\
HD 141937 b&       9.7 &      19.2&       33.5 & 2.6 &      1.5&      0.41 &       no&      no \\
%NORTH HD 33564 b&       9.1 &      18.64&       21.0 &       1.1 &      0.34 &       no&      no \\ 
HD 74156 c&       8.0 &      21.9&       64.6 &   3.7 &    3.4 &      0.43 &       no&      no \\ 
HD 39091 b&       10.3 &      18.5&       18.3 &  3.8 &     3.3&      0.61 &       no&      yes \\
HD 81040 b&       6.9 &      21.3&       32.6 &    4.2 &   1.9&      0.53&       no&      no \\
HD 106270 b&       11.0&      21.9&       84.9&  4.3 &     4.3 &      0.40&       no&      no \\
%NORTH HD 22781 b&       13.65 &      19.52&       31.8 &       1.2 &      0.82 &       no&      no \\
%NORTH 14 Her b&       4.64 &      22.08&       18.1&       2.8&      0.37&       no&      no \\
HD 111232 b&       6.8 &      21.7&       29.0&   5.2 &    2.0&      0.20&       no&      no \\ 
HD 142 c&       5.3 &      22.1&       20.6&         5.9 &  6.8&      0.21 &       no&      no \\ 
HD 106515 A b&       9.6 &      21.0&    35.2&  6.0 &     4.6&      0.57&       no&      yes \\
HIP 5158 c&       15.0 &      20.4&       45.0&    6.0 &   7.7&      0.14 &       no&      yes \\
HD 219077 b&       10.4 &      21.2&       29.4 &  8.9 &     6.2 &      0.77&       no&      yes \\
HD 62509 b&       2.9 &      22.4\tablenotemark{b}&       10.3 &  5.0\tablenotemark{c} &     1.7&     0.02&       no&      no \\
GJ 676 A b&       4.9 &      21.6&       16.5 &   5.0\tablenotemark{c}  &  1.8&      0.33&       no&      yes \\
%HD 196067 b&       6.9 &      22.45&       44.3&       5.0&      0.66 &       no&      no \\
\enddata
\tablenotetext{a}{Flag indicating whether object is also detected in the M band within the given detection limits.}
\tablenotetext{b}{L band brightness likely underestimated as first order estimates suggests $T_{eff,model} < T_{eq}$ (see text).}
\tablenotetext{c}{Age assumed for analysis as no age was given in exoplanet.eu database (see text).}
\end{deluxetable*}

\subsection{Detecting small planets}
One of the mid- to long-term goals of exoplanet research is certainly the direct detection and characterization of rocky - and potentially habitable - planets. It is useful to consider what ELTs might be able to deliver in this context. In a first step, we provide some first-order estimates of what parameter space in terms of planet size, temperature and host star properties E-ELT/METIS will be able to probe, depending on the observing wavelength. In a second step, we carry out an updated version of the Monte-Carlo experiment first presented by \citet{crossfield2013} to quantify, how many planets we can expect to detect based on the occurrence rate of planets found by the \emph{Kepler} mission.

\subsubsection{Small planet parameter space probed by E-ELT/METIS}
The following first order estimates provide some interesting insights about the prospects of imaging small planets with E-ELT/METIS. We consider three planetary sizes (1 R$_\earth$, 2 R$_\earth$, and 3 R$_\earth$) and five different effective temperatures for these planets approximated as blackbody emission (255 K, 300 K, 400 K, 500 K, 600 K). Varying the distance between Earth and these planets, we compute the flux density received at Earth in different wavebands. As a benchmark test it is useful to recall that Earth seen from 10 pc distance emits roughly 0.4 $\mu$Jy around 10.5 micron assuming black-body emission coming from its surface \citep{desmarais2002}.

In order to assess if certain types of planets can be directly imaged with METIS, we further need to take into account the sensitivity limits in each filter and also the IWA achievable at each observing wavelengths (Table~\ref{metis_table}). As described above, we assume that the BGL can be achieved at $2 \lambda_{{\rm cen}}/D$ and we take this separation as IWA. Finally, we assume that the planet's effective temperature corresponds to its equilibrium temperature  $T_{eq}$, which depends on the luminosity of the star, the planet's Bond albedo $A_B$, and the separation from the host star $r_p$: 
$$ T_{eq}=\Bigg[\frac{L_*(1-A_B)}{4\pi\sigma}\Bigg]^{1/4}\Bigg(\frac{1}{2 r_p}\Bigg)^{1/2}\quad .$$ 
Using the Earth as reference case with $T_{eq}=255$ K, $r_p = 1$ AU, $L_* = 1$ L$_\sun$, and $M_* = 1$ M$_\sun$, and recalling $T_{eq} \propto 1/\sqrt{r}$, we can estimate the planet-star separation that corresponds to different $T_{eq}$. For example, around the Sun, $T_{eq}=500$ K corresponds to a separation of 0.26 AU. In oder to be able to consider different spectral types for the host star, we approximate $T_{eq} \propto L_*^{1/4}$ and   $L_* \propto M_*^4$, which leads to $T_{eq} \propto M_*$. Hence, for an F star with a mass of 1.5 M$_\sun$ we find $T_{eq}=383$ K at 1 AU and, correspondingly, $T_{eq}=255$ K at 2.25 AU. Knowing the IWA of our instrument for a given wavelength we can now directly compare its value with the projected separation of any given planet-host star combination. We considered five host star types defined by their mass (0.5 M$_\sun$ (M-type), 0.75 M$_\sun$ (K-type), 1.0 M$_\sun$ (G-type), 1.5 M$_\sun$ (F-type), 2.0 M$_\sun$ (A-type)). 

\begin{figure*}
\centering
\epsscale{1.1}
\plotone{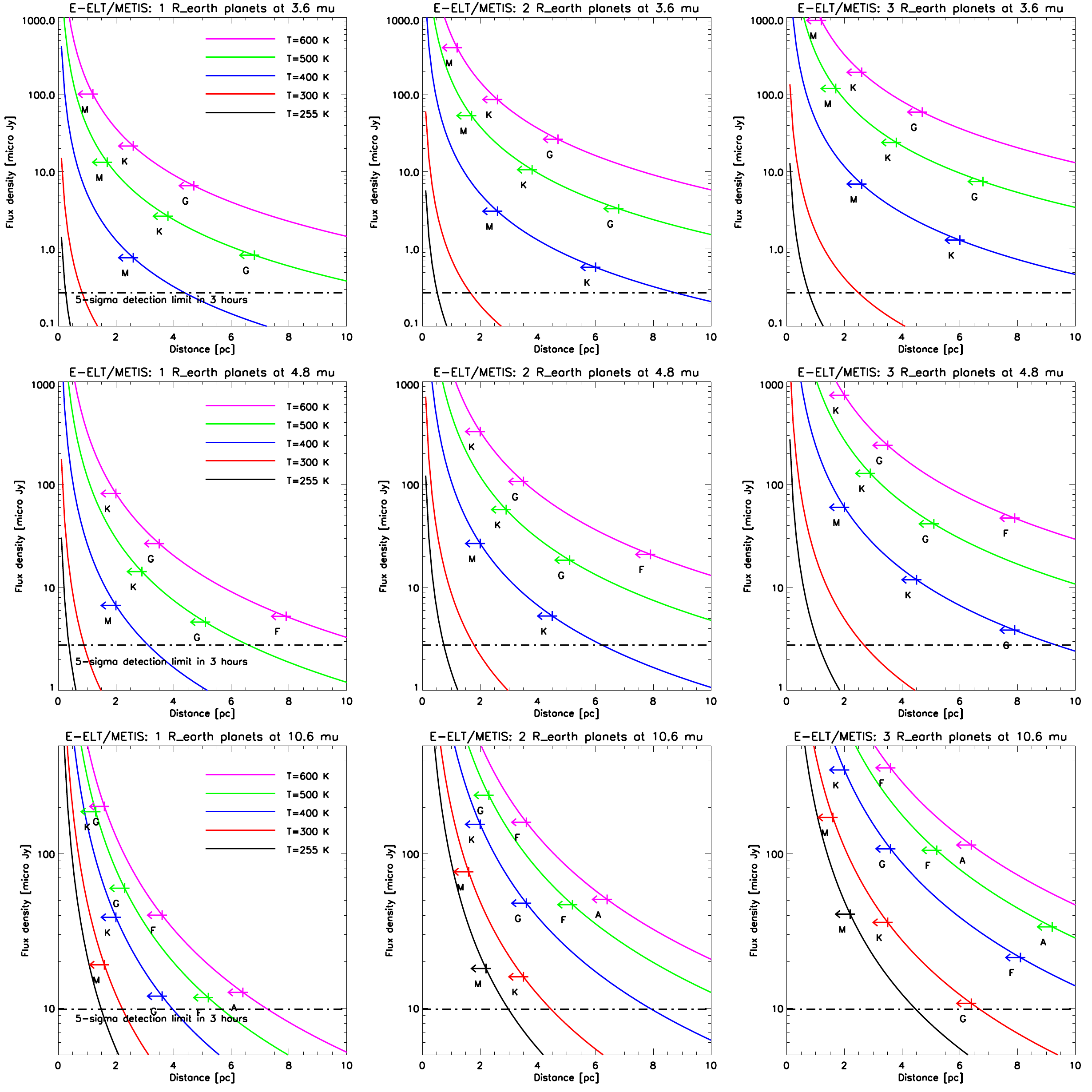}
\caption{First order estimate of the parameter space of small planets that METIS can probe at different wavelengths. All plots show the flux density of different kinds of small planets as a function of their distance from the Sun and are organized as follows: Analyses for the L band (top row),  M band (middle row), and N band (bottom row); planets with 1 R$_\earth$ (left column), 2 R$_\earth$ (middle column), and 3 R$_\earth$ (right column). The different colors correspond to different blackbody temperatures of the  planets (see legends in the left column). The arrows and the letters below them indicate out to what distance a planet with a given size and temperature can be detected around a star with a certain mass, i.e., spectral type (see text). At distances greater than this limit the assumed IWA is insufficient to spatially resolve the star-planet system in the given sensitivity limits. The dash-dotted lines denote the 5-$\sigma$ detection limit in 3 hours of telescope time for each filter (see, Table\ref{metis_table}).
\label{fig.2}}
\end{figure*}

Figure~\ref{fig.2} shows the results. The basic trends are intuitively clear: The hotter, the bigger and the more nearby the planet, the more flux is received at Earth. As hotter planets have to be closer to their host star, depending on the host star's spectral type, we reach the IWA of the telescope where we can no longer spatially separate the planet from the star at a certain distance from the Earth. While hotter planets emit more flux, the space volume we can probe, where we still spatially resolve them from their stars, is much smaller than for cooler planets. 

Looking at some planet-star combinations more specifically, we see two key results: (1) the coolest planets in our analyses ($T_{eq}=255$ -- 300 K) would only be detectable in the N band around the very nearest stars and neither in the L nor M band. (2) For all hotter planets ($T_{eq}=400$ -- 600 K) the L band is the best wavelength range for planet detections as for any planet-star combination the space volume probed is larger than in the other bands.  

\subsubsection{Monte-Carlo simulation of small planet detections with E-ELT/METIS}
To complement the analysis described above, we determined the population of planets that would be accessible to METIS observations using empirically constrained estimates of short-period planet frequency as a function of stellar type, planet radius and orbital period. To do this we used the Monte-Carlo approach outlined in detail in \cite{crossfield2013}. Briefly described, this analysis simulates thousands of plausible extrasolar planetary systems using measurements of planet frequency from the \emph{Kepler} satellite \citep{howard2012} and estimates the likelihood of finding a detectable planet by comparing the planets' predicted blackbody fluxes to analytic estimates of high-contrast performance \citep{guyon2005} and detection limits. Unlike in Section 3.1., here the planet's thermal emission does not come from residual heat of formation, but from absorbed and reprocessed starlight. Contributions from reflected starlight to the observed planet fluxes is also taken into account. For this, the planet's Bond albedo and geometric albedo are assumed to be wavelength independent and are randomly drawn from a uniform distribution between 0.0 -- 0.4 \citep{crossfield2013}.

\begin{figure}
\centering
\plotone{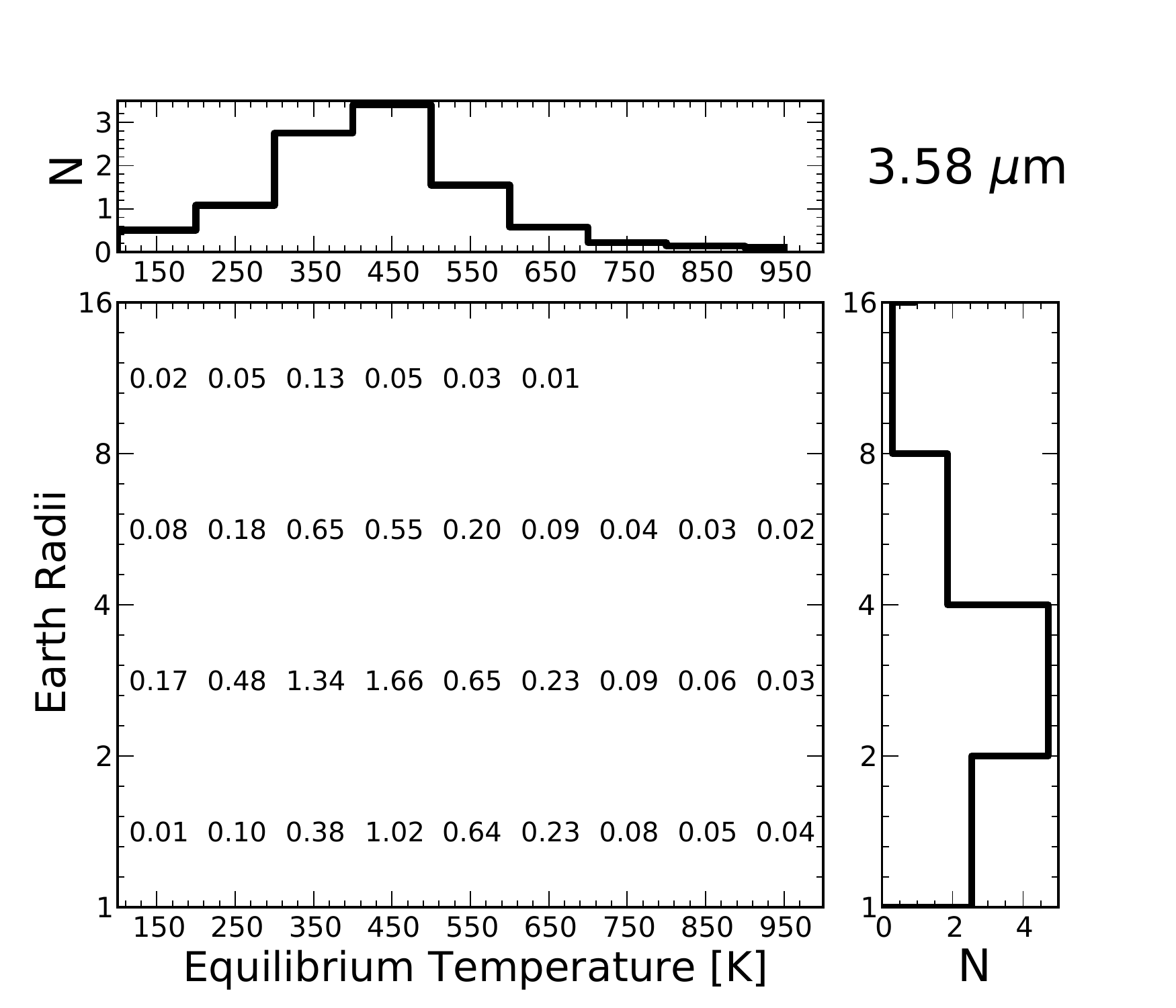}
\plotone{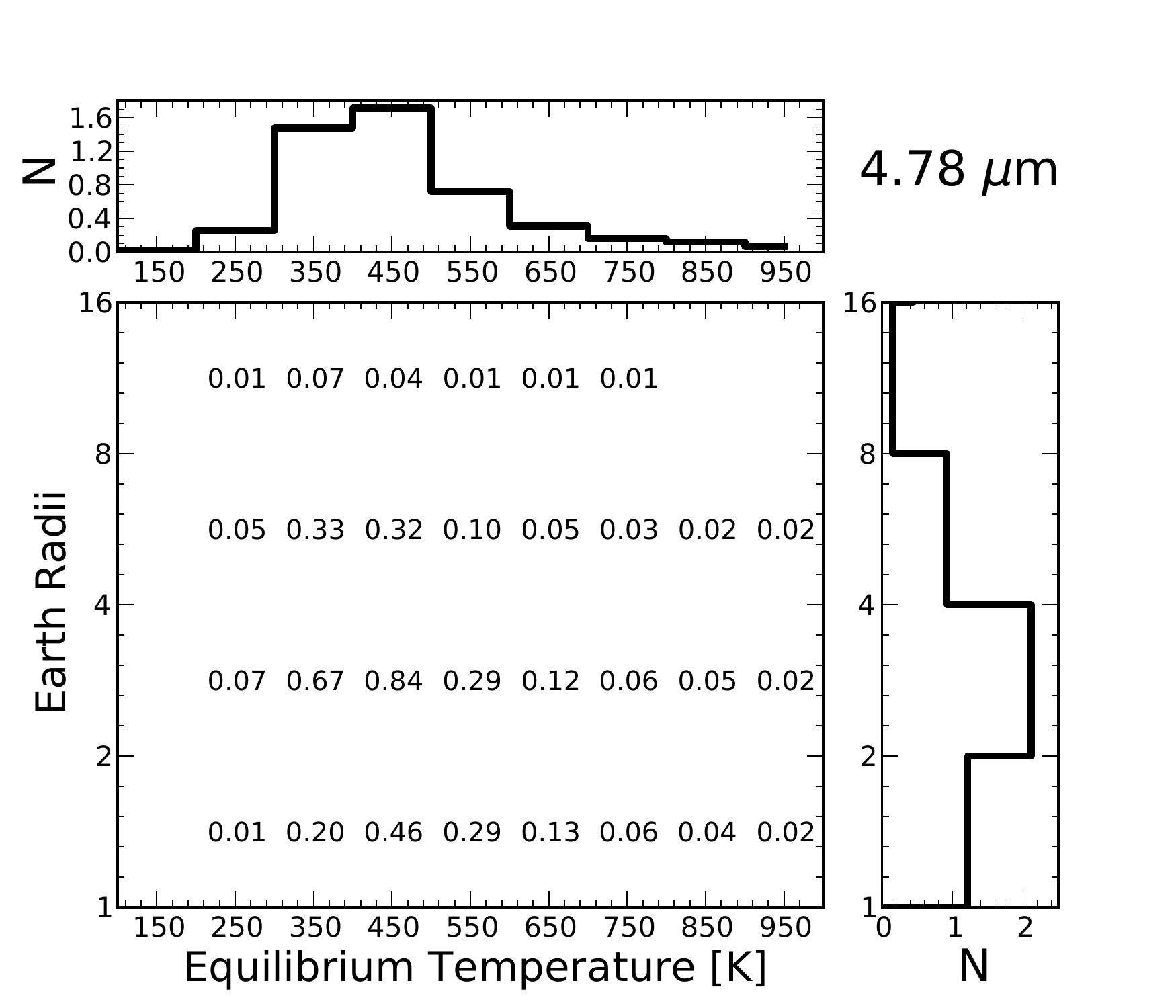}
\plotone{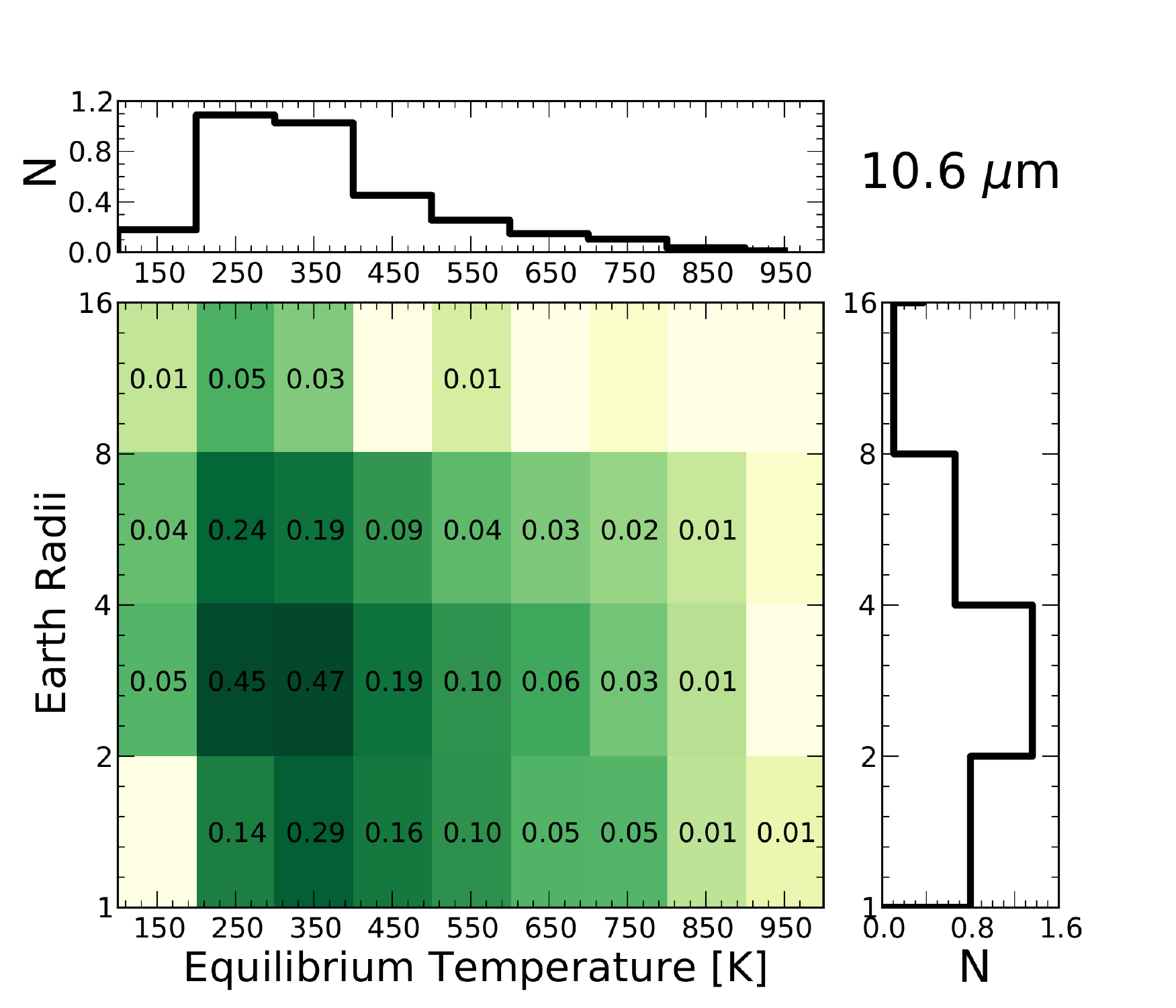}
\caption{2-D probability distributions for the detection of small planets using E-ELT/METIS. From top to bottom the panels show the distributions for detections in the L, M and N band, respectively.  \label{fig.3}}
\end{figure}

Compared to the original study done by \citet{crossfield2013}, which had the goal of comparing the expected detection yield of different wavelength regimes, instrument performances and  telescope sizes, the present study is different in a number of aspects. \citet{crossfield2013} focused primarily on contrast performance and did not explicitly include constraints coming from sensitivity estimates. Ground-based thermal infrared imaging has, however, quite severe sensitivity constraints coming from the high background emission at these wavelengths. The present analysis takes sensitivity limits explicitly into account (Table~\ref{metis_table}). Also, while \citet{crossfield2013} was a comprehensive but partly generic case study, the present analysis is for a specific telescope and a specific instrument. The corresponding sensitivity estimates were derived using an instrument simulator (including, e.g., a model PSF, sky background noise, throughput, relevant telescope parameters). Furthermore, in light of the results presented in the previous section, we extended our analysis beyond 8 pc \citep[as in][]{crossfield2013} to include all dwarf stars with $K<7$ mag and $d<20$~pc. We removed close binaries, gathered stellar photometry and parallaxes from the literature \citep[][and SIMBAD]{perryman1997, monet2003, cutri2003,zacharias2012}, adopted stellar radii and effective temperatures based on interferometric measurements of similar stars \citep{boyajian2012a,boyajian2012b,boyajian2013}, and assigned stellar masses using the V-band relation of \cite{henry1993}. Selecting only objects with declination $<$30$^\circ$, this final target list includes 246 objects; 24 of these host already known planets or planet candidates, most of which do not pass the detection threshold that we impose in our analysis, as they are either too close to the star and/or too faint. Finally, the present analysis determines how many planets are detectable in more than one filter (see below) and provides a concrete list of stars for which we summarize the detection probabilities per filter.

This new Monte-Carlo analysis reveals that $\approx$10 small planets within 15 pc should be detected in at least one of the L, M, or N bands. Roughly 5 objects could be observed in both L and M band, and a small number ($\sim$2) might be observable in a combination of N and L and/or M. The results are summarized in Figure~\ref{fig.3}, where we show the 2-D probability distributions (planetary radius vs. equilibrium temperature) separately for the L, M, and N band. Roughly 25\% of the planets have radii of 1--2 R$_{\earth}$. The rest has radii $> 2 R_{\earth}$ and is increasingly likely to host a substantial gaseous envelope \citep[cf.][]{marcy2014}. The expected $T_{eq}$ of the smaller planets is $\sim 100$~K higher than for the larger planets. This is a selection effect: In our simulations, larger planets ($\gtrsim4 R_{\earth}$) are seen mainly in reflected starlight (even in M band); smaller planets, however, must emit relatively more thermal radiation to climb above the sensitivity threshold, and so thermal radiation comprises up to $\sim50$\% of their observed flux. For the L and M band the most likely range of equilibrium temperature is 300 -- 500 K, while, statistically, in the N band a couple of planets in the 200 -- 400 K range should be found. 

For the results shown in Table~\ref{rocky_planets} we changed the perspective and analyzed which stars in our sample are the best targets for planet searches. We only list objects where the probability of detecting a planet - regardless of size or temperature - is at least 10\% in one of the observational bands. Table~\ref{rocky_planets} emphasizes that, according to our simulations, the L band is the best wavelength range to search for planets, but it also shows that - based on the \emph{Kepler} planet occurrence statistics - for some stars there is a fair chance to detect planets in more than one band. 

\section{Discussion}
The analyses presented in the previous sections rely on some assumptions and led to some results that warrant further discussion.

Concerning the assumed sensitivity limits it is obvious that these are preliminary and are probably subject to change in the course of the METIS project. However, the values represent the current state of knowledge. Similarly, the exact filter profiles are not yet defined for METIS. This leads to some uncertainties when we compare the predicted fluxes for the RV-detected gas giant planets to the METIS detection limits because the model predictions were computed for a different filter set. However, this effect is expected to be rather small, especially compared to the uncertainties in the models themselves. Assuming that METIS will achieve background-limited performance down to an IWA as small as $2 \lambda_{{\rm cen}}/D$ is certainly a challenging goal, both from a technical and a data processing point of view. Until now, in some cases, the background limit was only reached at $\sim 5 \lambda_{{\rm cen}}/D$ \citep[e.g.,][]{kenworthy2010}. The coming years will show whether the continuous progress in coronagraphic and data processing techniques is sufficient to enable such a demanding, but scientifically crucial, performance at the E-ELT. Finally, our analyses did not take into account that, depending on the final instrument design and coronagraphs used, additional observing time is required to achieve the detection limits assumed here in a full 360$^\circ$ circle around the target stars. Pupil plane coronagraphs will alter the general throughout of the instrument and do not necessarily create a centro-symmetric high-contrast region around the star for separations $>2 \lambda_{{\rm cen}}/D$ \citep{kenworthy2010,carlotti2013}. For focal plane coronagraphs additional overhead might be created by regular switching between the target star and a reference sky location for sky subtraction. 

\begin{deluxetable}{llccccc}
\tablecaption{Summary of Monte-Carlo results for specific nearby stars. All stars shown have a probability for planet detection with E-ELT/METIS of at least 10\% in one of the bands. The last three columns show the detection probability in the L, M and N band, respectively. 
\label{rocky_planets}}           % title of Table
\tablewidth{\linewidth}
\tablehead{
\colhead{Name}  & \colhead{Catalog} & \colhead{Spec.}   & \colhead{Dist.} &  \colhead{$p_L$} & \colhead{$p_M$} & \colhead{$p_N$} \\
\colhead{ }  & \colhead{name}   & \colhead{Type}   & \colhead{[pc]} &  \colhead{ } & \colhead{ } & \colhead{ } }
\startdata
        alpha Cen B*  &  HD 128621 & K1V & 1.3  & 0.59 & 0.65 &  0.74 \\
         alpha Cen A  &   HD 128620 &    G2V &  1.3 &  0.51 &  0.63 & 0.67 \\
         epsilon Eri*    &   	 HD 22049 &  	K2V  & 3.2  & 0.47 & 0.32  & 0.32 \\
       epsilon Ind A   &  HD 209100 &    K4V &  3.6  & 0.46 & 0.26  & 0.14 \\
             tau Cet*  &  	 HD 10700 &   G8.5V &  3.7 &  0.34 &  0.29 & 0.18 \\
         Proxima Cen  &    HIP 70890 & M5.5V  & 1.3 &  0.33 & --  & --\\
            Gl 166 A    & HD 26965 A&  K0.5V &  5.0 &  0.32 & 0.17  &--\\
           delta Pav    &  	 HD 190248 &  G8IV &  6.1 &  0.32 & 0.24 &  0.12 \\
           Procyon A  &  HD 61421  &  F5IV-V &  3.5 &  0.31 & 0.31 & 0.28 \\
              Gl 887    &   HD 217987 & M0.5V  & 3.3 &  0.28 & --  &--\\
              GJ 139*   &    HD 20794 & G8V &  6.0 &  0.27 & 0.15  &--\\
              Gl 825     & HD 202560&    K7V &  3.9 &  0.26 & --  &--\\
            beta Hyi     &  HD 2151&   G0V  & 7.5  & 0.22 & 0.12  &--\\
            LTT 2364    &    HD 38393 &  F6V &  9.0 &  0.22 & 0.13  &--\\
      Barnard's Star    &    HIP 87937 & M4V &  1.8 &  0.21 &  -- &--\\
             zet Tuc   &   HD 1581&   F9.5V &  8.6 &  0.17 & --  &--\\
            Gl 570 A    & 	 HD 131977 &    K4V  & 5.8 &  0.16 & --  &--\\
             HR 4523*    &   HD 102365 &  G2V &  9.2  & 0.16 & --  &--\\
             gam Pav    &  	 HD 203608 &   F9V  & 9.2 &  0.16 & --  &--\\
             LHS 348    &   HD 114710 &   G0V &  9.2 &  0.16 & --  &--\\
            LHS 2465    &   HD 102870 &   F9V & 10.9 &  0.15 & --  &--\\
           chi01 Ori     &   	 HD 39587 &  G0V &  8.7 &  0.15 & --  &--\\
             iot Peg    &    	 HD 210027 &  F5V & 11.8 &  0.14 & --  &--\\
            36 Oph C   &   HD 156026 &  K5V &  5.9  & 0.13 & --  &--\\
             gam Ser    &  HD 142860 &   F6IV & 11.1 &  0.13 & --  &--\\
             107 Psc     &  	 HD 10476 &  K0V  & 7.5 &  0.12 & --  &--\\
            Ross 154   &     HIP 92403 & M3.5V &  3.0 &  0.11 &  -- &--\\
            Sirius A   &   HD 48915 &  A0.5V &  2.6 &  0.11 & 0.11  & 0.10 \\
               1 Eri     &   HD 17206 &  F7V & 14.0 &  0.11 &  -- &--\\
             61 Vir*     &  HD 115617 &   G7V &  8.5 &  0.10 &  -- &--\\
\enddata
\tablenotetext{*}{Stars with known or suggested exoplanets: alpha Cen B \citep{dumusque2012}, epsilon Eri \citep{hatzes2000}, tau Cet \citep{tuomi2013}, GJ 139 
\citep{pepe2011}, HR 4523 \citep{tinney2011}, 61 Vir \citep{vogt2010}. }
\end{deluxetable}

For the RV-detected planets it is clear that some of the orbit parameters listed in the exoplanet.eu database may be refined by future observations or that for some systems different groups obtained slightly different results. However, the selection and detection criteria we applied were rather conservative (e.g., minimum mass, $T_{eq}<T_{eff,model}$) so that detection biases introduced by uncertain orbit parameter should not have a significant effect. Also, we did not take into account any possible contribution from reflected starlight to the emission coming from the planets. This contribution is, however, indeed in most cases negligible as the planet-star flux ratio is on average $\sim 10^{-6}$ only considering the planets' thermal emission, while the flux ratio in reflected light is typically at least 2 orders of magnitude smaller. More importantly we need to emphasize that the possible METIS target sample might significantly change in the coming years. With increasing time baselines of RV planet surveys, additional long period gas giant planets can expected to be found in the future, some of which yielding better detection probabilities than the objects listed here. In addition, the \emph{GAIA} spacecraft will reveal additional long period gas giant planets using high-precision astrometric measurements. Also here we can expect additional high-priority targets for direct imaging follow-up studies with E-ELT/METIS. 

The assumptions for the Monte-Carlo simulations of nearby small planets and their impact on the results have been discussed in length in \citet{crossfield2013}. One of the key assumptions is that the \emph{Kepler} results for the innermost $\sim$0.25 AU follow a flat distribution for wider orbits in logarithmic period space. However, this assumption is consistent with the recent analysis of four full years of Kepler data \citep{petigura2013}. Once a planet candidate has been imaged around a nearby star, follow-up observations within a few months will not only be able to confirm common proper motion of the planet and its host star, but they allow also for a robust determination of the planet's orbit and hence of the received stellar insolation as a function of orbital phase. An assessment of whether any of the small planets accessible to METIS are potentially habitable is clearly beyond the scope of this paper. To be sure, a broader wavelengths coverage is certainly required for an in-depth analysis of atmospheric features. However, it becomes more and more clear that life can exist over a broad temperature range and a current upper limit appears to be $\sim$400 K \citep[e.g.,][and references therein]{seager2013}. In addition, the basic concept of the 'habitable zone' around a star is undergoing major revisions, and of particular interest for the analyses presented here it seems that, under certain atmospheric conditions, the inner edge of the habitable zone around Solar type stars extends much closer in - and hence to higher equilibrium temperatures - than originally thought \citep{zsom2013}. To push the characterization aspect a bit further, a possible next steps could be to replace the assumed black-body curves with more realistic atmospheric models with different compositions and resulting albedos. Not only will this help us to further refine the detection probabilities for different planet types, but we can also quantify to what degree the different spectroscopy modes of METIS and other future instruments could be used to further characterize the small planets that we can expect to find.  

\section{Conclusions}
We explored two exoplanet science cases for the future 3 -- 10 micron instrument METIS planned for the E-ELT. One of the key assumptions in our analyses is that METIS will be background-limited at IWAs as small as $2 \lambda_{{\rm cen}}/D$, which is certainly technically challenging. We showed that already there is an interesting and sizable sample of RV-detected gas giant planets that METIS will be able to image directly within a reasonable amount of observing time. More than 20 objects covering a wide range of planetary masses and host star spectral types will be detectable in the L band, and half of those objects will also be seen in the M band given the detection limits we assume. With the continuation of RV searches for long period planets and with the advent of the \emph{GAIA} astrometry mission, many more, and perhaps better suited, targets will be added to the list in the future. Studying those objects in a grander sample with METIS will allow us for the first time to test and refine atmospheric models for planets where the distance, and the object's mass, orbit and luminosity are empirically determined. 

In addition to detecting cool gas giant planets, we showed that E-ELT/METIS could be the first instrument that might image a small and potentially rocky planet around one of the nearest stars. Based on the \emph{Kepler} statistics METIS might detect $\approx$10 planets within 15 pc in at least one band (L, M or N). The L band offers the broadest discovery space for all planet-star combinations we analyzed. Roughly 5 objects should be observable in both L and M band, and for a couple of objects we might get a detection in N, L and/or M. Statistically speaking, most of these planets are expected to have equilibrium temperatures between 300 -- 500 K, and at least a quarter of them will have a radius between 1 -- 2 R$_{\earth}$. Hence, if those planets do exist around some of the nearest stars, METIS might reveal an object with potentially habitable conditions. 
\newline\newline
{\it Acknowledements:} We thank the Journal for inviting us to contribute this paper to this special issue on exoplanets and we thank the 2 referees for useful comments and suggestions. SPQ thanks A. Zsom for interesting discussions and M. Reggiani for providing an interpolation tool to determine the luminosity of gas giant planets as a function of their age and mass. This work made use of the on-line SIMBAD database, operated at CDS, Strasbourg, France.

\end{document}